\documentclass{amsart}

\newtheorem{theorem}{Theorem}
\newtheorem{corollary}[theorem]{Corollary}
\newtheorem{thm}{Theorem}[section]

\newtheorem{lemma}[thm]{Lemma}

\theoremstyle{definition}

\theoremstyle{remark}
\newtheorem{remark}[thm]{Remark}

\numberwithin{equation}{section}

\def \eqskip { \vspace*{3mm} }

\newcommand{\RM}{\mathbb{R}}

\newcommand{\eps}{\varepsilon}

\newcommand{\ka}{{\kappa}}

\newcommand{\ups}{{\phi}}

\newcommand{\disps}{\displaystyle}

\begin{document}

\title{On minimal eigenvalues of Schr\"{o}dinger
operators on manifolds}
\author{Pedro Freitas
\thanks{Partially supported by FCT, Portugal}
}
%
%
\address{Departamento de Matem\'{a}tica,
Instituto Superior T\'{e}cnico, Av.Rovisco Pais, 1049-001 Lisboa,
Portugal.}
\email{pfreitas@math.ist.utl.pt}
\date{\today}
%

\begin{abstract}
We consider the problem of minimizing the eigenvalues of the
Schr\"{o}d\-inger operator $H=-\Delta+\alpha F(\ka)$ ($\alpha>0$) on
a compact $n-$manifold subject to the restriction that $\ka$ has a
given fixed average $\ka_{0}$.

In the one--dimensional case our results imply in particular that for
$F(\ka)=\ka^{2}$ the constant potential fails to minimize the
principal eigenvalue for $\alpha>\alpha_{c}=\mu_{1}/(4\ka_{0}^{2})$, 
where $\mu_{1}$ is the first nonzero eigenvalue of $-\Delta$. This
complements a result by Exner, Harrell and Loss, showing that
the critical value where the circle stops being a minimizer for a
class of Schr\"{o}dinger operators penalized by curvature is given by
$\alpha_{c}$. Furthermore, we show that the value of $\mu_{1}/4$ remains
the infimum for all $\alpha>\alpha_{c}$. Using these results, we
obtain a sharp lower bound for the principal eigenvalue for a general
potential.

In higher dimensions we prove a (weak) local version of these results
for a general class of potentials $F(\ka)$, and then show that globally
the infimum for the first and also for higher eigenvalues is actually given
by the corresponding eigenvalues of the Laplace--Beltrami operator and
is never attained.

%
\end{abstract}

\maketitle
\section{Introduction}
\label{intro}
In the last years there has been a great interest in the study of
optimal properties of eigenvalues of Schr\"{o}dinger operators of the
form $H= -\Delta + V$ defined on compact manifolds, when some
restrictions are imposed on the potential $V$. Some of these problems
are related to several physical phenomena such as motion by mean
curvature, electrical properties of nanoscale structures, etc (see,
for instance,~\cite{a,ahs,ei,ehl,hl,ke} and the references therein).

In~\cite{ehl}, the authors considered the case of potentials depending
on the curvature $\ka$ and studied the problem of minimizing the first
eigenvalue of the operator $H = -d^{2}/ds^{2} +
\alpha\ka^{2}$ defined on a closed planar curve with length one. They
proved that for $0<\alpha<1/4$ the circle is the unique minimizer,
while for $\alpha>1$ this is no longer the case, leaving open the
question of the value of $\alpha$ where the transition takes place,
and also what happens after this critical value.

More generally, one might consider an operator $H$ defined on a
compact $n-$manifold $(M,g)$ by $H=-\Delta+\alpha F(\ka)$ and with
eigenvalues $\lambda_{0}<\lambda_{1}\leq\ldots$, and study the
problem of determining
\[
\Lambda_{j}(\alpha) = \inf_{\ka\in K}\lambda_{j}(\ka), \; j=0,1,\ldots
\]
where
\[
K=\left\{\ka\in\mathcal{C}(M;\RM): \frac{\disps 1}{\disps
|M|}\int_{M}\ka dv_{g}=\ka_{0}
\right\}.
\]
In particular, we are interested in knowing whether or not there
exists a critical value of $\alpha$, say $\alpha_{c}$, where the
constant potential stops being a global minimizer for the first 
eigenvalue. In this paper we
show that in the one--dimensional case studied in~\cite{ehl} this
critical value is in fact equal to $1/4$, and that for $\alpha$ larger
than $\alpha_{c}$ the infimum is identically equal to $\pi^{2}$ and
is not attained. The first part of this result is a consequence of a
more general result which provides an upper bound for $\alpha_{c}$
holding in any dimension. Furthermore, we show that for potentials of
the form $\ka = \ka_{0}+\eps q$ where $q$ has zero average, this bound
is in fact precise for sufficiently small values of $\eps$, in the
sense that for $\alpha$ smaller than the bound, the constant potential
$\ka_{0}$ gives a smaller eigenvalue than $\ka$, while for larger
values of $\alpha$ this is not always the case.

These results could lead us to expect that results similar to those in 
one dimension would also hold in higher dimensions, that is,
that there would exist a nontrivial interval $(0,\alpha_{c})$ where
the constant potential was the unique minimizer. However, it turns out
that for dimensions higher than the first there exist potentials
satisfying the given restrictions and which make the principal
eigenvalue as close to zero as desired. Thus, we see that in this case
the constant potential is never a global minimizer. It remains an open
question if it is a local minimizer. A similar statement also holds
for higher eigenvalues and for minimizations subject to other types of
integral restrictions -- see Theorem~\ref{theo3} and the remarks that
follow it. The reason for this different behaviour in dimensions
higher than the first is directly related to the fact that in this 
case, given
a manifold $M$ and a geodesic ball $B_{\delta}$ of radius $\delta$
centred at a point $x_{0}$ in $M$, the Dirichlet eigenvalues of the
Laplacian in $\Omega_{\delta} = M\setminus B_{\delta}$ converge to
those of the Laplacian in $M$ as $\delta$ approaches zero -- for a
more precise statement of this property see Section~\ref{highd}
and~\cite{cf}.

Finally, we point out that the results in one dimension enable us to
obtain a lower bound for the principal eigenvalue in the case of a
general potential (Corollary~\ref{cor2}) which, of course, corresponds
also to the first eigenvalue of Hill's equation. Note that one of the
motivations behind the study of the minimization of eigenvalues when
the potential is subject to integral restrictions was precisely to
obtain lower bounds for eigenvalues -- see~\cite{ke}.

\section{Notation and general local results\label{secgenloc}}
Let $(M,g)$ be a compact Riemannian $n-$manifold with metric $g$ and
let $-\Delta$ denote the Laplace--Beltrami operator defined on $M$
with eigenvalues $0=\mu_{0}<\mu_{1}\leq\ldots$ repeated according to
their multiplicity. Denote the corresponding orthonormal (with respect
to the $L^{2}(M)$ inner product induced by the Riemannian measure
$v_{g}$) system of eigenfunctions by
$\left\{v_{j}\right\}_{j=0}^{\infty}$. Consider now the operator
defined on $M$ by $H=-\Delta +\alpha F(\ka)$, where
\[
\frac{\disps 1}{\disps |M|}\int_{M} \ka dv_{g}=\ka_{0},
\]
and $F:\RM\to\RM$ is assumed to be of class $\mathcal{C}^{3}$ in a
neighbourhood of $\ka_{0}$.

The main result in this section is then the following

\begin{theorem}\label{th}
Assume that $F'(\ka_{0})\neq0$ and define
\[
\alpha^{*} =
\frac{\mu_{1}\disps F''(\ka_{0})}{\disps 2[F'(\ka_{0})]^{2}}.
\]
Then, if $q$ is a continuous real valued
function with zero average and not identically zero, we have that for
$\ka=\ka_{0}+\eps q$ with sufficiently small $\eps$ (depending on $q$
and $\alpha$), the principal eigenvalue $\lambda_{0}$ of
$H$ satisfies
\[
\lambda_{0}(\ka)>\alpha F(\ka_{0}), \mbox{ if } \; 0<\alpha<\alpha^{*},
\]
while for $\alpha>\alpha^{*}$ there exist functions $q$ as above for
which
\[
\lambda_{0}(\ka)<\alpha F(\ka_{0}).
\]
\end{theorem}
\begin{proof}
Consider the Schr\"{o}dinger operator defined on $M$ by
\[
H_{\eps} = -\Delta + \alpha F(\ka_{0}+\eps q).
\]
Since $M$ is compact, the spectrum of $H_{\eps}$ is discrete and its
first (simple) eigenvalue and the corresponding (normalized)
eigenfunction are analytic functions of the (real) parameter
$\eps$~\cite{kat}. We thus expand $\lambda_{0}$ and the corresponding
eigenfunction $u$ as a power series of $\eps$ around zero:
\[
\begin{array}{l}
\lambda_{0} = \ell_{0}+\ell_{1}\eps+\ell_{2}\eps^{2}+\ldots\eqskip\\
u =  \ups_{0}+\ups_{1}\eps+\ups_{2}\eps^{2}+\ldots.
\end{array}
\]
On the other hand, we also have that $F(\ka_{0}+\eps q) =
f_{0}+f_{1}q\eps +f_{2}q^{2}\eps^{2}+o(\eps^{2})$, where
\[
f_{0} = F(\ka_{0}),\ f_{1} = F'(\ka_{0}), \mbox{ and }\; f_{2} =
\frac{\disps 1}{\disps 2}F''(\ka_{0}).
\]

Substituting these expressions in the equation giving the eigenvalues
for $H_{\eps}$ we obtain, equating like powers in $\eps$,
\[
\begin{array}{lc}
\eps^{0}: & -\Delta \ups_{0} + \alpha f_{0}\ups_{0} = \ell_{0}\ups_{0}\\
\eps^{1}: & -\Delta \ups_{1} + \alpha f_{0}\ups_{1} + \alpha f_{1}q \ups_{0} =
\ell_{0}\ups_{1}+\ell_{1}\ups_{0}\\
\eps^{2}: & -\Delta \ups_{2} + \alpha f_{0}\ups_{2} + \alpha f_{1}q\ups_{1} +
\alpha f_{2}q^{2}\ups_{0} = \ell_{0}\ups_{2}+\ell_{1}\ups_{1}+\ell_{2}\ups_{0}.
\end{array}
\]
From the first equation it follows that $\ell_{0}=\alpha f_{0}$ and
that $\ups_{0}$ is constant, which we take to be one. Substituting
this in the equation for $\eps^{1}$ and integrating over $M$ gives
that $\ell_{1}$ vanishes and $\ups_{1}$ satisfies
\begin{equation}
\label{eqp1}
-\Delta \ups_{1} = -\alpha f_{1}q.
\end{equation}
Substituting now this in the last equation gives that $\ups_{2}$
satisfies
\[
-\Delta \ups_{2} = -\alpha f_{2}q^{2}-\alpha f_{1}q\ups_{1} + \ell_{2}.
\]
Again integrating over $M$ gives
\begin{equation}
\label{l2exp}
\ell_{2} = \frac{\disps \alpha f_{2}}{\disps |M|}\int_{M}q^{2}dv_{g}+
\frac{\disps \alpha f_{1}}{\disps |M|}\int_{M}q\ups_{1}dv_{g}.
\end{equation}
Taking squares on both sides of~(\ref{eqp1}) we get
$\left[\Delta(\ups_{1})\right]^{2} = \alpha^{2} f_{1}^{2}q^{2}$. On the other
hand, multiplying the same equation by $\ups_{1}$ and integrating over
$M$ gives that
\[
\alpha f_{1}\int_{M}q\ups_{1}dv_{g}=-\int_{M}|\nabla
\ups_{1}|^{2}dv_{g}.
\]
Substituting these two expressions into~(\ref{l2exp}) we finally
obtain
\[
\ell_{2} = \frac{\disps f_{2}}{\disps \alpha f_{1}^{2}|M|}\int_{M}
(\Delta \ups_{1})^{2}dv_{g}- \frac{\disps 1}{\disps
|M|}\int_{M}|\nabla \ups_{1}|^{2}dv_{g},
\]
and it follows from Lemma~\ref{lemb} below that $\ell_{2}$ is always
positive for $\alpha<\alpha^{*}$.

To give an example of a function $q$ for which $\ell_{2}$ becomes
negative when $\alpha>\alpha^{*}$ it is sufficient to take $q=v_{1}$.
We obtain from~(\ref{eqp1}) that in this case
\[
\ups_{1}=c-\frac{\disps \alpha}{\disps \mu_{1}} f_{1} v_{1},
\]
where $c$ is an arbitrary constant. Substituting this into the
expression for $\lambda_{2}$ yields
\[
\ell_{2} = \frac{\disps \alpha}{\disps |M|}\left(f_{2}-
\frac{\disps \alpha}
{\disps \mu_{1}} f_{1}^{2}\right),
\]
which is negative for $\alpha>\alpha^{*}$.
\end{proof}

An obvious consequence of this result is that for all $F$ of the form
above there exists a value of $\alpha$, say $\alpha^{**}$ such that
for $\alpha>\alpha^{**}$ the constant potential is not a minimizer of
the first eigenvalue.

In the case where $F'$ is allowed to vanish, it is also clear that if
$\ka_{0}=\ka_{0}^{*}$ is a (local) minimizer
(resp. maximizer) of  $F$, it follows that, for positive values of 
$\alpha$, $\ka(x)\equiv\ka_{0}^{*}$ will be a (local) minimizer 
(resp. maximizer). This is the case, for instance, when 
$F(\ka)=\ka^{2}$ and $\ka_{0}=0$, where obviously $\ka=0$ is a global 
minimizer for all $\alpha$.

The result needed to prove that $\ell_{2}>0$ for $\alpha<\alpha^{*}$
is neither new nor difficult, but a specific reference could not be
found in the literature and so, for the sake of completeness, we
provide a proof here.
\begin{lemma} \label{lemb}The functional
\[
I_{\alpha}(u) = \int_{M}\alpha(\Delta u)^{2}-|\nabla u|^{2}dv_{g}
\]
is nonnegative for $\alpha\geq 1/\mu_{1}$.
\end{lemma}
\begin{proof}
The spectral problem corresponding to $I_{\alpha}$ is
\begin{equation}
\label{auxeig}
\alpha\Delta^{2}u+\Delta u = \gamma u,
\end{equation}
which has discrete spectrum $\gamma_{0}\leq\gamma_{1}\leq\ldots$. We
will prove that if $\alpha> 1/\mu_{1}$ then $\gamma_{j}\geq0$ for all
$j=0,1,\ldots$. To this end rewrite~(\ref{auxeig}) as
\[
\Delta(\alpha\Delta u+u) = \gamma u.
\]
It is not difficult to see that $u$ is an eigenfunction if and only if
\[
\alpha\Delta u + u = \beta v_{j}
\]
for some real number $\beta$ different from zero. For
$\alpha>1/\mu_{1}$ the operator $\alpha \Delta + I$ is invertible and
thus this last equation has one and only one solution given by
$u=\beta v_{j}/(1-\alpha\mu_{j})$. Substituting this
into~(\ref{auxeig}) gives $\gamma = (\alpha\mu_{j}-1)\mu_{j}$ from
which the result follows.
\end{proof}

\section{The one--dimensional case}\label{sec1d}

In this section we consider the particular case studied in~\cite{ehl}
with $F(\ka) = \ka^{2}$, and for which
\[
\alpha^{*} = \frac{\disps \mu_{1}}{\disps 4\ka_{0}^{2}}.
\]
As a consequence of Theorem~\ref{th} and the results in~\cite{ehl} we
have the following
\begin{theorem}\label{cor1}
In the one dimensional case and for $F$ as above, $\alpha_{c}=\alpha^{*}$.
Furthermore, for
$\alpha>\alpha_{c}$, $\Lambda_{0}(\alpha)\equiv \mu_{1}/4$.
\end{theorem}
\begin{proof}
It only remains to show the result for $\alpha$ larger than
$\alpha_{c}$. Clearly in this case $\Lambda_{0}(\alpha)\geq\mu_{1}/4$.
Consider now the family of potentials given by
\[
\ka_{\delta}(s) =\left\{
\begin{array}{ll}
\ka_{0}/\delta, & 0<s<\delta,\\
0, & \delta<s<\ell.
\end{array}
\right.
\]
Note that although $\ka_{\delta}$ is not continuous on the circle, it
can be approximated by continuous functions without affecting our
results. For this family of potentials we obtain the functional
\[
J_{\delta}(u) = \int_{0}^{\ell}\left[u'\right]^{2}ds +
\frac{\disps \alpha\ka_{0}^{2}}{\disps \delta^{2}}\int_{0}^{\delta}u^{2}ds,
\]
where $u$ is normalized. We now take $u(s) = \sqrt{2}\sin(\pi s/\ell)$ to
obtain
\[
J_{\delta}(u)=\frac{\disps \mu_{1}}{\disps 4}+\frac{\disps 2\alpha\ka_{0}^{2}}
{\disps \delta^{2}}\int_{0}^{\delta}\sin^{2}(\frac{\disps \pi 
s}{\disps \ell})ds,
\]
and since
\[
\lim_{\delta\to 0^{+}}\frac{\disps\int_{0}^{\delta}\sin^{2}(\pi s)ds}
{\disps \delta^{2}} = 0,
\]
it follows that $J_{\delta}$ can be made to be arbitrarily close to $\mu_{1}/4$.
\end{proof}
\begin{remark}{\rm
Clearly for $\alpha>\alpha_{c}$ the infimum is not attained, as was
conjectured in~\cite{ehl}.}
\end{remark}

A simple consequence of Theorem~\ref{cor1} is a lower bound for the
principal eigenvalue of the Schr\"{o}dinger operator on the circle.
\begin{corollary}\label{cor2}
Consider the operator $H=-d^{2}/dx^{2} + V(x)$ defined on $(0,L)$ with
periodic boundary conditions, and define
\[
V_{m}=\inf_{x}V(x) \mbox{ and } I =
\frac{\disps 1}{\disps L}\int_{0}^{L}\left[V(x)-V_{m}\right]^{1/2}dx.
\]
Then
\[
\lambda_{0} \geq\left\{
\begin{array}{ll}
V_{m}+I^{2}, & \mbox{ if } I\leq \frac{\disps \pi}{\disps L}\\ V_{m}+
\frac{\disps \pi^{2}}{\disps L}, & \mbox{ if } I>\frac{\disps \pi}{\disps L},
\end{array}
\right.
\]
with equality for $I<\pi/L$ if and only if $V$ is constant.
\end{corollary}
\begin{proof}
The first inequality follows directly by writing the eigenvalue
problem as $-u''+(V-V_{m})u = (\lambda-V_{m})u$ and applying the
previous corollary with $\ka = (V-V_{m})/\alpha$. The second part is a
consequence of the fact that for $\alpha$ larger than $\alpha_{c}$ the
principal eigenvalue must be larger than $\alpha_{c}\ka_{0}^{2}$.
\end{proof}
\begin{remark} {\rm
It follows from Theorem~\ref{cor1} that the given inequalities are
sharp in both cases.}
\end{remark}

\section{Higher dimensions}\label{highd}

In~\cite{ehl}, the proof of the fact that for $\alpha$ smaller than
$\alpha^{*}$ the constant potential is the unique global minimizer of
$\Lambda_{0}$ relied on a result that is not available in higher
dimensions. Namely, while in one dimension we have that
\[
\int_{S^{1}}\left[(u-u_{m})'\right]^{2}ds\geq\frac{\disps \mu_{1}}{\disps 4}
\int_{S^{1}}(u-u_{m})^{2}ds,
\]
where $u_{m}$ is the minimum of $u$ in $S^{1}$, from the results
in~\cite{cf} it is known that there is no similar result in higher
dimensions. More precisely, if we impose that a function $f$ be zero
at a finite number of points of a compact manifold with dimension
greater than or equal to two, then there is no relation of the form
above with a positive constant on the right--hand side. This suggests
that an argument similar to that used in the proof of
Theorem~\ref{cor1} can now be used for all positive values of
$\alpha$, and not just for $\alpha$ larger than $\alpha^{*}$. This is
indeed the case, and we have the following
\begin{theorem}\label{theo3}
Assume that $F(0)$ is a global minimum of $F$. Then, for $n$ greater
than one, $\Lambda_{j}(\alpha)\equiv \mu_{j}-F(0)$ for all positive
$\alpha$ and $j=0,1,\ldots$.
\end{theorem}
\begin{proof}
Fix a point $x_{0}$ in $M$ and denote by $B_{\delta}$ the geodesic
ball centred at $x_{0}$ with radius $\delta$. Let now
$\Omega_{\delta}=M\setminus\overline{B_{\delta}}$ and define the
potential
\[
\ka_{\delta}(x) = \left\{
\begin{array}{cl}
\frac{\disps \ka_{0}}{\disps |B_{\delta}|}, & \mbox{ if }
x\in B_{\delta}(x_{0}),\eqskip\\ 0, & \mbox{ if } x\in
\Omega_{\delta}(x_{0})
\end{array}
\right.
\]
(As before, this is discontinuous but can be approximated by
continuous functions without changing the results.) By subtracting
$F(0)$ on both sides of the equation for the eigenvalues, we can,
without loss of generality, take $F(0)$ to be zero. We are thus lead
to the functional
\[
J_{\delta}(u) = \int_{M}|\nabla u|^{2}dv_{g} +
\alpha F\left(\frac{\disps \ka_{0}}{\disps |B_{\delta}|}\right)
\int_{B_{\delta}} u^{2}dv_{g}.
\]

Consider now the auxiliary eigenvalue problem defined by
\[
\left\{
\begin{array}{cl}
-\Delta w = \mu w, & \; x\in\Omega_{\delta}\eqskip\\
w = 0, & \; x\in \partial\Omega_{\delta},
\end{array}
\right.
\]
and denote its eigenvalues by
$0<\mu_{0}(\delta)<\mu_{1}(\delta)\leq\ldots$, with corresponding
normalized eigenfunctions $v_{j\delta}$. From the results in~\cite{cf}
we have that
\[
\lim_{\delta\to0^{+}}\mu_{j}(\delta) = \mu_{j}, \;\; j=0,\ldots.
\]

We now build test functions $u_{j\delta}$, $j=0,\ldots$ defined by
\[
u_{j\delta}(x) = \left\{
\begin{array}{cl}
v_{j,\delta}(x), & x\in\Omega_{\delta}\eqskip\\ 0, & x\in
B_{\delta},
\end{array}
\right.
\]
for which
\[
J_{\delta}(u_{j\delta}) = \int_{\Omega_{\delta}}|\nabla
v_{j,\delta}|^{2} dv_{g},
\]
and, by the result from~\cite{cf} mentioned above, this converges to
$\mu_{j}$, $j=0,\ldots,$ as $\delta$ goes to zero. Finally, note that
for each $\delta$ the set $\left\{u_{j\delta}\right\}_{j=0}^{\infty}$
satisfies the necessary orthogonality conditions, since this is the
case for $\left\{v_{j\delta}\right\}_{j=0}^{\infty}$
\end{proof}
A similar result will also hold in other cases, such as manifolds with
boundary with Dirichlet or Neumann boundary conditions, for instance.

\section{Concluding remarks}

As was pointed out in~\cite{ehl} for the one--dimensional case, it is
not difficult to see that for negative $\alpha$ the constant potential
still {\it maximizes} the principal eigenvalue. It is also possible to
show that in this case there is no lower bound on this eigenvalue, in
the sense that there exist potentials $\ka$ with fixed average
$\ka_{0}$ for which this eigenvalue can be made as large (in absolute
value) as desired. It is not completely clear what happens to the
supremum of the first eigenvalue for positive values of $\alpha$.

Regarding higher dimensions, it was shown that integral restrictions
of this and similar type actually impose no restrictions at all as far
as minimization is concerned, in the sense that it is possible to
approximate the eigenvalues of the Laplacian as much as desired by
potentials satisfying the given restrictions. Although we have seen
that in this case the constant potential is never a global minimizer
for positive $\alpha$, the results in Section~\ref{secgenloc} raise
the question of whether or not it is a local minimizer for
$\alpha<\alpha^{*}$.

We end by remarking that similar results to those in
Sections~\ref{secgenloc} and~\ref{sec1d} also hold in the case of 
manifolds with boundary and Neumann boundary conditions.

\subsection*{Acknowledgements} This work was carried out while I was visiting
the Department of Mathematics of the Royal Institute of Technology in
Stockholm, Sweden. I would like to thank the people there and, in
particular, Ari Laptev, for their hospitality.

%
%

\end{document}